\documentclass[prl,showkeys,twocolumn]{revtex4}
\usepackage{graphicx,amssymb,amsmath}
\usepackage{hyperref}
\begin{document}

\title{Modeling the relation between income and commuting distance}

\author{Giulia Carra}
\affiliation{Institut de Physique Th\'{e}orique, CEA, CNRS-URA 2306, F-91191,
Gif-sur-Yvette, France}
\author{Ismir Mulalic}
\affiliation{DTU Denmark}
\affiliation{Kraks Fond - Institute for Urban Economic Research}
\author{Mogens Fosgerau}
\affiliation{DTU Denmark}
\author{Marc Barthelemy}
\email{marc.barthelemy@cea.fr}
\affiliation{Institut de Physique Th\'{e}orique, CEA, CNRS-URA 2306, F-91191,
Gif-sur-Yvette, France}
\affiliation{Centre d'Analyse et de Math\'ematique Sociales, (CNRS/EHESS) 190-198, avenue
de France, 75244 Paris Cedex 13, France}

\begin{abstract}

We discuss the distribution of commuting distances and its relation to
income. Using data from Denmark, the UK, and the US, we show that the
commuting distance is (i) broadly distributed with a slow decaying tail that
can be fitted by a power law with exponent $\gamma \approx 3$ and (ii) an
average growing slowly as a power law with an exponent less than one that
depends on the country considered. The classical theory for job search is
based on the idea that workers evaluate the wage of potential jobs as they
arrive sequentially through time, and extending this model with space, we
obtain predictions that are strongly contradicted by our empirical findings.
We propose an alternative model that is based on the idea that workers
evaluate potential jobs based on a quality aspect and that workers search
for jobs sequentially across space. We also assume that the density of
potential jobs depends on the skills of the worker and decreases with the
wage. The predicted distribution of commuting distances decays as $1/r^{3}$
and is independent of the distribution of the quality of jobs. We find our
alternative model to be in agreement with our data. This type of approach
opens new perspectives for the modeling of mobility.

\end{abstract}


\maketitle

\section{Introduction}

Cities are growing and the majority of individuals in the world now
live in urban areas~\cite{UN_Nat}. Understanding what governs the
evolution and the organization of urban systems is thus of primary
interest for policy makers and planners. The availability of large
scale data about almost all aspects of cities has opened the
possibility of a new interdisciplinary science of cities with solid
foundations~\cite{Batty:2013}. In particular, understanding mobility
patterns is a central problem in this field and is related to the
labor market, a fundamental area of interest in economics, where the
choice of work and residential locations determines the commuting. We
focus here on a part of this area, namely the job search process which
has a direct impact on the spatial distribution of commuting
trips. The seminal contributions on job search theory in economics
\cite{LippmanMccall:1976,McCall:1970, Stigler:1961} rely on the
central assumption that individuals choose among different job offers
that arrive sequentially in time, by maximizing their expected
discounted net wage, while waiting to accept a job offer is
costly. These are clearly very strong assumptions that should be
tested against empirical data.

Surprisingly, the standard model of job search \cite{McCall:1970} does
not integrate space (some labour market studies do take into
account, see for example \cite{Zenou:2009}). We introduce here a
spatial component in this model and derive the consequences for the
distribution of the commuting distance. In particular, we show that
the basic McCall model \cite{McCall:1970} does not explain some
fundamental statistical features observed in empirical data. We
therefore propose a new stochastic model that does not rely on the
assumption of optimal control in search through time, but instead on
the idea that workers search through space, accepting an offer if it
has a certain level of `quality'. The quality of a job is random and
unobserved by the researcher and it may integrate any number of
quality aspects specific to each individual. We find excellent
agreement between this new model and empirical data for Denmark, the
UK, and the US.

Beyond the prediction of the distribution of commuting distances and
their relation with income, our model provides a search-based
microfoundation for models of spatial patterns that can be found in
the mobility literature \cite{Simini:2012}. More generally, we
question here the relevance of optimal control theory as the main
framework to explain mobility and the behavior of living
organisms. Optimal control theory is a mathematical optimization
method used to find the policies that optimize the outcome of a given
process. This method has been applied to many different problems in
areas such as biology, economics and finance, ecology, and
management~\cite{appli_1, appli_2, appli_3, appli_4}. Here, we propose an alternative
framework to study human or animal behavior, closer to theories
developed about foraging \cite{foraging} and in which actions are
taken not on the basis of an optimal strategy but on the first
opportunity that is judged to be good enough.

In the first part of this paper, we present an empirical analysis of the
distribution of commuting distances for Denmark, the UK, and the US,
exploring how the average commuting distance scales with income. In a second
theoretical section, we derive the probability distribution for the
commuting distance from the spatial extension of the standard job search
model. We compare this theoretical prediction with our empirical results,
and show that the standard theoretical framework is not in agreement with
data. We then propose a new stochastic model which does not rely on the
optimal strategy assumption, and where workers evaluate potential jobs
sequentially across space and based on a quality aspect. We then show that
this new model is in excellent agreement with our data.

\section{Empirical results}

In this section, we investigate the distribution of commuting distances and
its relation to individual income using datasets for three different
countries: Denmark, the UK~\cite{UKdata}, and the US~\cite{USdata}. These
datasets are produced by national agencies and national household surveys
(see Materials and Methods for details) and record the commuting distance
and the income range at the individual level. The datasets cover the whole
of each country and take into account all transportation modes. For the UK
the data is for the years 2002-2012, for the US three different years are
available (1995, 2001, 2009), and for Denmark we have access to 10 years
(2001-2010).

\subsection{The average commuting distance}

We first focus on the simplest quantity, the average commuting distance, and
how it varies with income. The results for the three countries studied here
are shown in Fig.~1 (left column). The basic equilibrium models of urban
economics \cite{Alonso:1964,Muth:1969,Brueckner:2000} predict, within a
single city, that workers with higher incomes will have longer commuting
distances. This prediction is confirmed for Denmark and the UK, while no
particular trend can be detected for the US.


For Denmark, we observe an increasing range and a saturation at large income
values, while for the UK we observe a plateau at low income values. In the
range where the increase is observed we can fit the data by a power law of
the form 
\begin{equation}
\overline{r}(Y)\sim Y^{\beta }
\end{equation}%
where $Y$ is the individual income and where the exponent $\beta $ depends
on the country considered. For the US, the fit gives an exponent $\beta
\approx 0$ indicating that there is no clear trend. For the UK, the plateau
around the commuting distance value $\overline{r}\approx 5$ miles occurs in
the low income range $[10^{2},10^{4}]$ (GBP/year). The fit on UK data for
incomes higher than $5,000$ GBP (for all modes and all years) gives an
exponent value $\beta \approx 0.5$ (in the range $[0.53,0.66]$ when
considering different years). In contrast, we observe for the Danish data a
strong dependence with a large exponent of order $0.8$ for yearly incomes
larger than $250,000$ DKK and smaller than $500,000$ DKK (for lower incomes
we observe a small plateau). Depending on the year considered, the exponent $%
\beta $ varies in this case in the range $[0.61,0.88]$.

\begin{figure}[ht!]
\includegraphics[width=1\columnwidth]{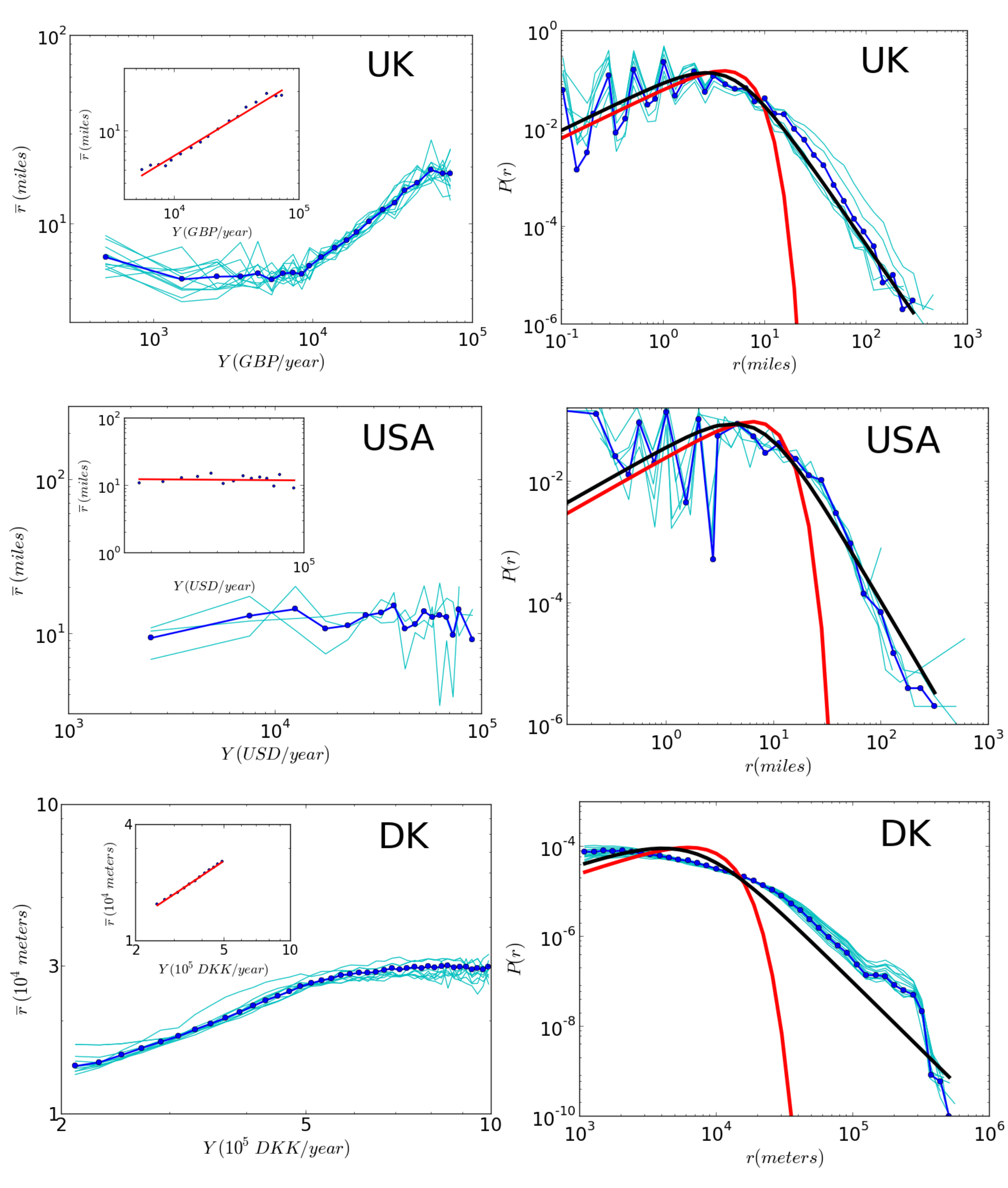}
\caption{\textbf{Left column: Average commuting distance versus income for
different years.} In dark blue, the commuting distance is averaged over all
years. (Top) UK data. This loglog plot displays a plateau for small values
of income followed by a regime, when fitted by a power law (see inset),
gives an exponent $\protect\beta \approx 0.5$ ($[0.53,0.66]$). In the inset
the average commuting distance is averaged over all years and the power law
fit gives an exponent $\protect\beta = 0.58$. (Middle) US data. In this
loglog plot we do not observe an income dependence. Indeed, a power law fit
gives an exponent $\protect\beta \approx 0$. (Bottom) Danish data. The power
law fit on the commuting distance averaged over all years (in the inset)
gives an exponent $\protect\beta = 0.77$. \textbf{Right column: Commuting
distance distribution for different income classes.} The probability
distribution is shown for different income classes. In dark blue we show the
distribution for a particular value of the income for which fits have been
performed. In red, we show the one parameter fit with the analytical
function predicted by the extended McCall model (Eq.~\protect\ref{eq:Pr}),
and in blue, the one parameter fit with the analytical function predicted by
the closest opportunity model (Eq.~\protect\ref{eq:pri}). (Top) UK data
(averaged over all available years). (Middle) US data (averaged over all
available years). (Bottom) Danish data (all years give the same result and
we choose here to show the year $2008$). In all cases we observe that the
tail predicted by the extended McCall model (Eq.~\protect\ref{eq:Pr}) decays
too quickly and cannot fit the data for long distances. In contrast, the
closest opportunity model is in excellent agreement with empirical
observations. }
\label{fig:1}
\end{figure}

\subsection{The distribution of commuting distance}

We now consider the full distribution of the commuting distance, shown in
Fig.~1 (right column) for different incomes for Denmark, the UK, and the US.
There are two important facts that we can extract from these empirical
observations. First, for all datasets studied here, the distribution is
broad. This means that the variation range of commuting distances is
extremely large. Indeed, we observe that with a non-negligible probability,
individuals in Denmark, UK, and US are commuting on distances of the order
of a few hundred kilometers. Second, the shape of the distribution and the
large distance behavior are remarkably similar among the different countries
we have studied here. These non-trivial features are very important as they
provide an opportunity to test for any model that aims to describe spatial
commuting patterns.

\section{Theoretical modeling}

The three datasets observed here display a slow increase of the average
commuting distance with income and, more importantly, a slowly decaying tail
for large distances. We would like to understand these two characteristics
theoretically. We begin with a discussion of the standard job search model
of economics \cite{LippmanMccall:1976,McCall:1970,Stigler:1961}, and compare
its predictions with our empirical observations. This will lead us to
propose another model, the `closest opportunity' model with predictions that
are in much better agreement with the data at hand.

\subsection{The spatial optimal job search model}


Optimal control theory is a well known mathematical optimization method used
to find policies that maximize the benefit of a given process. An example of
its application is the stopping problem~\cite{Chow:1971}, where one has to
choose the optimal time to take an action based on successive observations
of a random variable. Optimal control theory has been applied in many
different areas~\cite{Wald:1947, Bradt:1956, Shiryaev:1963, Haggstrom:1966,
Rasmussen:1979}, and to the job search problem~in economics \cite%
{LippmanMccall:1976,McCall:1970,Stigler:1961}. As a starting point, we will
here consider the important McCall model \cite{McCall:1970} that has been
used in many different forms and variants. We will study the implications of
the McCall model for the spatial distribution of distances between
residences and jobs depending on the income.

We begin by describing the McCall model in its simplest version. The job
search process is sequential in time. A worker who is unemployed at time $0$
reviews at every time step a random wage offer $w$ drawn from a distribution
with density $f$ (and cumulative $F$). At each time step, the worker can
either accept the current job offer and keep it forever, or she can pay a
waiting cost $c$ to discard the offer and wait for the next offer. The
worker's income $y_{t}$ at time $t$ will thus be $y_{t}=w$ if she accepts
the offer or $y_{t}=-c$ if she refuses it. The actual value of her total
returns is the discounted sum of her future payoffs 
\begin{equation*}
\sum_{t=0}^{\infty }\mu ^{t}y_{t}~,
\end{equation*}%
where the discount factor $\mu <1$ takes into account that the value of a
given amount of money is higher the earlier it is received. In this model,
with an offer $w$ at hand, the worker maximizes the expected value of her
total return $v(w)$ 
\begin{equation}
v(w)=\langle \sum_{t=0}^{\infty }\mu ^{t}y_{t}\rangle \;\;,
\end{equation}%
where the brackets denote the average over the offer distribution. The
classical way to solve this problem is to write the Bellman equation for
this stopping process which reads~\cite{Bellman:1957} 
\begin{equation}
v(w)=\max \left\{ \frac{w}{1-\mu },-c+\mu \int v(w^{\prime })f(w^{\prime })%
\mathrm{d}w^{\prime }\right\} ~.
\end{equation}%
This equation has a simple interpretation. The value of the current offer $%
v(w)$ is the maximum of two terms: the first term is the total return if the
current job offer is accepted, and the second term is the expected value of
rejecting the current offer and waiting for the next. In the latter case,
the worker pays the waiting cost $c$ and evaluates the expectation of the
value $v\left( w^{\prime }\right) $ of the next random offer $w^{\prime }$.
The optimal strategy that solves this equation is to accept the current
offer if it is larger than a \textit{reservation wage} $\tau $ and to refuse
it if it is lower. The reservation wage satisfies the equation 
\begin{equation}
\frac{\tau }{1-\mu }=-c+\frac{\mu }{1-\mu }\left[ \tau F(\tau )+\int_{\tau
}^{\infty }w^{\prime }f(w^{\prime })\mathrm{d}w^{\prime }\right] ~,
\end{equation}%
so that the worker is indifferent between accepting the job for which $%
w=\tau $ or waiting for another offer. By solving this equation, we obtain a
function $\tau $ that depends on the offer distribution. The probability $p$
of accepting an offer is then 
\begin{equation}
p=\int_{\tau }^{\infty }f(w)\mathrm{d}w~,  \label{eq:p}
\end{equation}%
and the number of trials $N$ before accepting a job offer thus follows a
geometric distribution 
\begin{equation}
P(N)=(1-p)^{N-1}p~.  \label{eq:PN}
\end{equation}

Space is absent in the McCall model and we will now extend it in the
simplest possible way. We assume now that the individual reviews the
job offers sequentially in the order of increasing distance from
home. The first offer reviewed is the closest to her residence, the
second one is the second closest and the $n^{th}$ time step
corresponds to the $n^{th}$ closest job to the seeker residence.  Each
random wage offer $w$ is still drawn from a distribution with density
$f$ (and cumulative $F$)and thus the probability that the individual
accepts an offer is still given by Eq.~\eqref{eq:p}. This means that
the worker, starting from home, will examine the offer and will choose
the first one that is above her reservation wage. We will also assume
that jobs are uniformly distributed in space with density $\rho $. If
a worker has accepted the $N^{th}$ offer, the probability that she has
moved a distance $r $ from its residence is given by a classical
result for the $N^{th}$ nearest neighbors in dimension $d=2$ for
uniformly distributed points ~\cite%
{Diggle:1983}
\begin{equation}
P(R=r|N)=\frac{2}{(N-1)!}\frac{1}{r}(\rho \pi r^{2})^{N}\mathrm{e}^{-\rho
\pi r^{2}}~.
\end{equation}%
The distribution of the commuting distance $R$ is then given by 
\begin{equation}
P(R=r)=\sum_{N\geq 1}P(r|N)P(N)
\end{equation}%
and since the distribution of $N$ is geometric (Eq.~\eqref{eq:PN}), we
obtain 
\begin{equation}
P(R=r)=2p\rho \pi r\mathrm{e}^{-p\rho \pi r^{2}}~.  \label{eq:Pr}
\end{equation}%
This distribution decreases as a gaussian over a scale of order $\sim 1/%
\sqrt{\rho p}$ where $1/\sqrt{\rho }$ corresponds to a typical interdistance
between different offers ($\tau $ and therefore $p$ depend on the income $Y$
and so does this distance too). We also note that the average commuting
distance decreases if the spatial density of opportunities $\rho $
increases. A decrease in the number of job openings during economic
downturns then leads to increasing commuting distances.

To test the consistency of these result with empirical data, we fit in
Fig.~\ref{fig:1} (right column) empirical data using the prediction
Eq.~\eqref{eq:Pr} of the extended McCall model. We observe that the
best (one parameter) fit is reasonable for the short distance regime
but is unable to reproduce the slow decay observed for large
distances. In addition, we have also considered another generalization
of the McCall model with transport costs, and showed that it also
cannot reproduce a slow decaying tail such as a power law (see the
general argument presented in the Materials and Methods section). It
thus seems that the McCall model is not consistent with our data. We
therefore seek an alternative model that does predict the empirical
findings just outlined. We will propose such a model in the next
section and compare its predictions with data.


\subsection{The closest opportunity model}

In this new model proposed here, we change three important assumptions of
the McCall model. First, we assume that workers evaluate offers sequentially
across space, whereas in the original McCall model the evaluation was performed
through time. Second, jobs are chosen based on some `quality' aspect that
could take into account many factors and not only on the wage (see for
instance \cite{Hornstein:2007,Hall:2013}). Finally, we change the framework
used to study human behavior, and the reservation wage of the McCall model,
which is the result of an optimal strategy, is replaced by a\textit{\
reservation quality} representing the minimal job quality that meets worker
expectations.

We still consider the problem of a worker who looks for a job starting
from her residence (that we assume to be located at $r=0$). Job offers
are uniformly distributed across space with density $\rho $. The density of jobs
$\rho$ relevant for the worker depends on the income level $Y$ and we assume that it is simply
\begin{equation}
\rho =\frac{\rho _{0}}{Y^{\alpha }}
\end{equation}%
such that higher income jobs are less dense than lower income
jobs. The exponent $\alpha $ depends on the country under
consideration and reflects many exogenous factors concerning job
offers at a certain income level \cite%
{Hornstein:2007,Hall:2013}. We remark that the job density $\rho $ is
the only parameter that discerns here different types of workers. We
also note that the framework introduced here for the income allows for many
generalizations to other quantities such as the skill level for example.

The McCall model assumes that jobs are primarily characterized by the wage
they offer. We depart from this and assume instead that each job is
characterized by a random `quality' $X$ that encodes many factors. The job
quality is distributed according to $f$ (with corresponding cumulative
distribution $F$) and job qualities are independent. We further assume that
a given worker has a \textit{reservation quality} value $\tau $ (in the same
spirit as the reservation wage), and she will keep expanding her search
radius until this threshold is met. We denote by $R$ the commuting distance
and its cumulative thus reads 
\begin{equation}
P(R\leq r|\tau )=P(X_{[0,r]}\geq \tau )=1-F(\tau )^{\rho \pi r^{2}}.
\end{equation}%
We now take into account that workers have different search costs and
different expectations for a future job, which leads them to have different
reservation qualities. We consider the reservation quality as random,
distributed according to a density $g(\tau )$, and obtain the cumulative
distribution of commute distances 
\begin{equation}
P(R\leq r)=\int g(\tau )P(R\leq r|\tau )\mathrm{d}\tau ,
\end{equation}%
with corresponding density 
\begin{eqnarray}
P(R=r) &=&\frac{dP(R\leq r)}{dr}  \notag \\
&=&-2\rho \pi r\int g(\tau )F(\tau )^{\rho \pi r^{2}}\log F(\tau )\mathrm{d}%
\tau .  \label{eq:1}
\end{eqnarray}%
The first term in this integral is the probability that a worker has
reservation quality $\tau $, the second term is the probability that all
offers are below $\tau $ in the disk of radius $r$, and the last term (the
logarithm) corresponds to the probability that at least one offer is above $%
\tau $ in the circular band $[r,r+\mathrm{d}r]$ (see Fig.~2 for a simple
illustration of this process). A simple and natural assumption for the
distribution of the reservation quality $\tau $ is that it is the same as
the distribution of job quality $F$: $g(\tau)\equiv f(\tau)$. Then Eq.~\eqref{eq:1} simplifies in a
remarkable way as follows 
\begin{eqnarray}
P\left( R=r\right)  &=&-2\rho \pi r\int f\left( \tau \right) F\left( \tau
\right) ^{\rho \pi r^{2}}\log \left( F\left( \tau \right) \right) \mathrm{d}%
\tau   \notag \\
&=&-2\rho \pi r\int_{0}^{1}x^{\rho \pi r^{2}}\log x\;\mathrm{d}x  \notag \\
&=&\frac{2\rho \pi r}{\left( 1+\rho \pi r^{2}\right) ^{2}}.
\end{eqnarray}%
Under these assumptions, the distribution of commuting distances does not
depend on the distribution of job quality, an effect that was already
observed in the specific case discussed in \cite{Simini:2012}, and the model
proposed here can then be considered as a microfoundation for this type of
process. This also means that we may generalize the interpretation of the
model: we may allow the distribution of job quality to be specific to each
worker, since this has no consequence for the distribution of commuting
distances.

\begin{figure}[ht!]
\includegraphics[width=0.5\columnwidth]{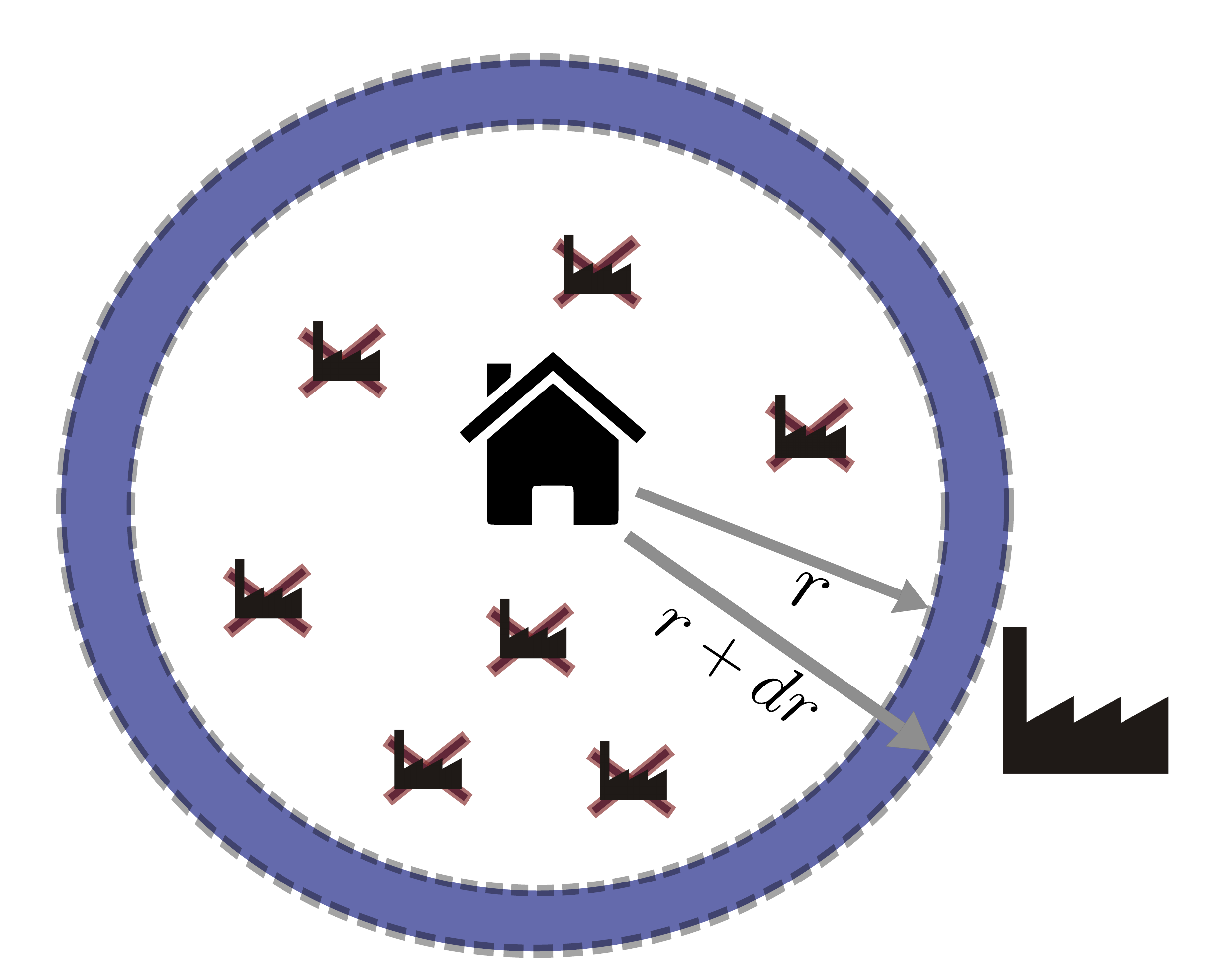}
\caption{Illustration of the argument leading to Eq. \eqref{eq:1}.}
\label{fig:2}
\end{figure}
In contrast to the McCall job-search model of the previous section that
displayed a rapid gaussian decaying tail, we observe here that the
distribution is slowly decaying as $P(R=r)\sim r^{-3}$ for large $r$. The
average commuting distance is easily computed within the closest opportunity
model and we find 
\begin{equation}
\overline{r}=\frac{1}{2}\sqrt{\frac{\pi }{\rho }}.
\end{equation}%
Replacing $\rho $ by $\rho _{0}/Y^{\alpha }$, we find that the distribution
of commute distance conditional on income is 
\begin{equation}
P\left( R=r|Y\right) =\frac{2\rho _{0}\pi rY^{\alpha }}{\left( Y^{\alpha
}+\rho _{0}\pi r^{2}\right) ^{2}}  \label{eq:pri}
\end{equation}%
and that the average commute distance is 
\begin{equation}
\overline{r}(I)=\frac{1}{2}\sqrt{\frac{\pi }{\rho _{0}}}Y^{\alpha /2},
\end{equation}%
which is a power law with exponent $\beta =\alpha /2$.

The theoretical result Eq.~\eqref{eq:pri} also implies a simple scaling that
can be checked empirically. Indeed, if we rescale the commuting distance by $%
Y^{\alpha/2}$, $u=r/\sqrt{Y^\alpha} $, all the curves for different incomes
should collapse on the unique curve that depends on only one parameter and
is given by 
\begin{equation}
P(u)=\frac{2\pi \rho_0 u}{(1+ \rho_0 \pi u^{2})^{2}}~.  \label{eq:Pu}
\end{equation}

In the next section, we evaluate these theoretical predictions against our
data.

\subsubsection{Comparison with empirical results}

The closest opportunity model predicts that the average commuting
distance varies with income as $\overline{r}\sim Y^{\alpha /2}$, where
$\alpha $ depends on the country considered. We will interpret our
empirical results in terms of this relationship. For the US, we
observe an exponent $\beta _{US}\approx 0$ indicating that the density
of jobs is independent from the skill level in the US. For the UK and
Denmark, we observe a non-zero exponent with $\beta _{UK}\approx 1/2$
for the UK and a larger value for Denmark $\beta _{DK}\approx
0.8$.
These results indicate that the density of jobs decreases with the
skill level, more in Denmark than in the UK. The observed difference
between the US and two European countries in the spatial density of
jobs at different income levels suggests a more general difference
between Europe and the US (for a discussion in equilibrium theory
about the spatial distribution of workers and skill levels, see for
example \cite{Brueckner:2002}). It is interesting to note that there seems
to be a correlation between the value of the exponent $\beta$ and the
size of the country. Further studies are however needed in order to
confirm this observation. 

The crucial prediction allowing us to distinguish between models is
the distribution of commuting distances and how it depends on
income. Indeed, for the simple spatial extension of the McCall model
presented here, the distribution of $r$ decreases very quickly
(Eq.~\eqref{eq:Pr}) and is not a broad distribution (extending the
McCall model with transport costs can lead to a broad distribution
such as a power law but this requires fine-tuning of parameters, see
the material and methods section). In sharp contrast, in the closest
opportunity model, we have a broad distribution of the form given by
Eq.~\eqref{eq:pri} and in Fig.~1 we display the one parameter fit with this form for
a given income category. The agreement with data is very good for the
UK and the US, but there are some discrepancies in the Danish case. It
seems that for this Danish case there are other heterogeneities that
are not taken into account in our model. In particular, Denmark is a
small country with a large proportion of the population living in
islands, imposing important constraints on commuting patterns.

An additional and very strong test of the validity of
Eq.~\eqref{eq:pri} is provided by the data collapse on the curve given
by Eq.~\eqref{eq:Pu}. In Fig.~3 (right column) we plot the rescaled
commuting distance distribution for different income categories and we
observe a very good collapse, except for the lower income category in
the UK for which the square root behavior is not applicable. We remark
that for the US $\beta =0$, which implies that the probability
distribution Eq.~\eqref{eq:pri} does not depend on the income category
so that the curves are automatically collapsed. We furthermore note
that the agreement between the data and the closest opportunity model
for Denmark is strongly reinforced by the data collapse predicted
by our model and observed in the data (shown in figure 3).

\begin{figure}[ht!]
\includegraphics[width=1\columnwidth]{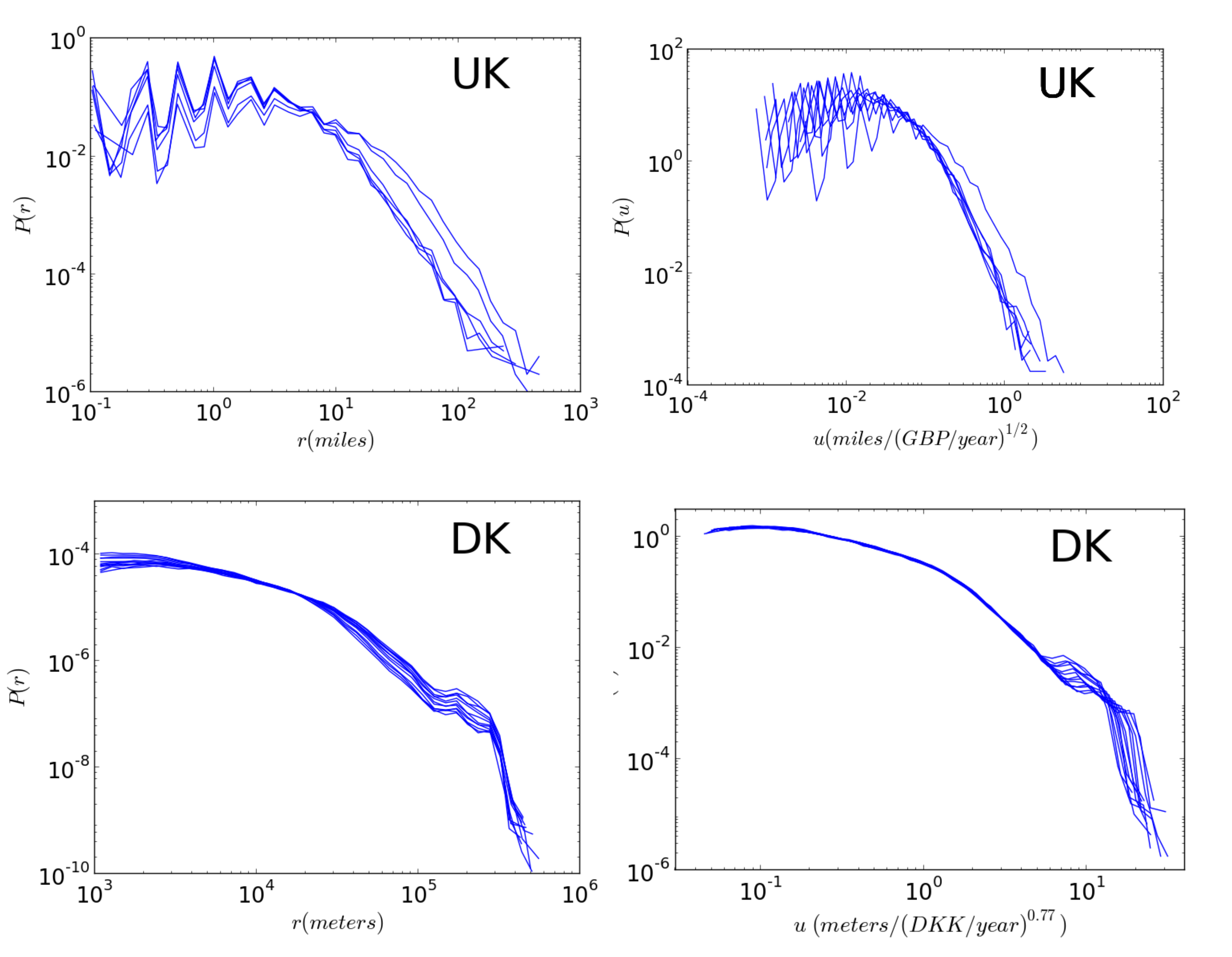}
\caption{ \textbf{Left column: Commuting distance distribution for different
income classes.} The probability distribution is shown for different income
classes. (Top) UK data (averaged over all available years). (Bottom) Danish
data (all years give the same result and we choose here to show the year $%
2008$). \textbf{Right column: rescaled probability distribution $P(u)$ for
different income classes.} We observe a very good data collapse for both UK
data (top), with $\protect\beta=0.5$ and averaged over all available years,
and the Danish data (bottom) for $\protect\beta=0.77$ and for the year $2008$%
.}
\label{fig:3}
\end{figure}
\section{Discussion and perspectives}

With the increasing availability of ever more precise and
comprehensive data we can test a number of predictions of models for
the urban structure and its processes. In this paper we predict the
distribution of commuting distances and discuss its relation with
income. We showed that the empirical data do not support the standard
McCall model (based on optimal control) for the job search
process. Instead, we have proposed a model based on the closest
opportunity that meets the expectation of each individual is able to
predict correctly the behavior of the average commuting distance with
income in terms of the density of jobs offers. More importantly, this
model is able to correctly predict the form of the commuting distance
distribution, its broad tail, and the data collapse predicted by its
form. 

Stated succinctly, previous models relied on the idea that workers
wait for a job that pays enough, while in the new closest opportunity
model, workers search space for a job that is good enough. Although
further studies on more countries are certainly needed, this
stochastic model provides a microscopic foundation for a large class
of mobility models and opens many interesting research directions in
modeling mobility while leading to testable predictions. More
generally, we proposed here an alternative framework to study human or
animal behavior, in which actions are taken not on the basis of an
optimal strategy but on the first opportunity that is good
enough. This framework would potentially find some applications in our
understanding of foraging for example and other applications in
ecology or finance where optimal control might be a too strong
assumption.



\section{Materials and Methods}

\subsection{Data description}

As we describe below, for both the US and the UK dataset a weighting
methodology has been developed in order to take into account non-responses,
undercoverage, multiple telephones in a household (for the US dataset) and
drops-off in the travel recording (for the UK dataset). This
methodology has been developed in order to make data trustable and
usable, but without any doubt, there is noise in the data (and probably self-reporting
errors too). One can indeed note that there is a bias for low income
values (for both for UK and USA data) which is very likely due to rounding.
However, this bias does not change the order of magnitude of the commuting
distance and thus does not substantially affect the results.

\subsubsection{UK data}

We used data from the UK National Travel Survey (NTS) for the years $%
2002-2012 $~\cite{UKdata}. Each year's sample has a size of $15,048$
addresses and was designed to provide a representative sample of households
in the UK. A weighting methodology was developed to adjust for
non-responses and drop-offs in the travel recording. Data collection is
obtained from face-to-face interviews and a seven day travel record of
individual daily travel activity.

We specifically exploit the \textit{individual} and the \textit{trip} files
of this dataset. The \textit{individual} file is used to determine the
income category of each individual (data provides $23$ income bands). The 
\textit{trip} file allows us to link individuals to their weekly commuting
trips for which we know the distance. In order to compute the average
commuting distance as a function of the income class, we first average the
commuting distance of each individual, including all transportation modes,
over the number of commuting trips undertaken during the week. We then
average these quantities over all individuals for each income category. When
we consider average values from these data, we do not distinguish between
different transportation modes or the geographical locations of the origin
and destination of the trip.

\subsubsection{US data}

We used data from the $1995$, $2001$ and $2009$ national household travel
survey (NHTS)~\cite{USdata}, a survey of the civilian, non-institutionalized
population of the United States. The NHTS datasets contain data for
respectively $42,033$, $26,032$,  and $150,147$ households (with approximately 40,000 add-on interviews
for the latest version). Weighting factors are
used in order to take into account nonresponses, undercoverage, and multiple
telephones in a household.

These datasets allow us to associate an income category to each worker (this
dataset indicates $18$ different income bins) and the one-way distance to
workplace. For the $2009$ NHTS, the personal income is not provided, in this
case we proxy personal income by the household income divided by the
household size.

\subsubsection{Danish data}

The Danish data are derived from annual administrative register data from
Statistics Denmark for the years $2001-2010$. We observe the full population
of workers, and for each year, we have information on the workers annual
income and their commuting distance. We used the post-tax income. Commuting
distances have been calculated using information on exact residence and
workplace addresses using the shortest route in between. Note that for these
data, no weighting methodology is required as we observe the full population
of workers in the country.

\subsection{Including transport cost in the McCall model}

We discuss here the general case for the McCall model where there is a
transport cost associated with distance. The distance from the home of
worker to a job offer is then a random variable $R$ having density $2\pi
\rho r$, which is independent of the wage $W$ associated with the job. In
order to link the probability of accepting a job to space, we assume a
linear transport cost $\delta R$ that is paid by the worker if she accepts a
job. Ultimately, she cares about the net wage $W-\delta R$. The optimal
strategy of the worker involves a reservation wage $\tau $ and the worker
accepts the first offer that offers a net wage $W-\delta R>\tau $. These
assumptions already imply that the commuting distance for the accepted job
satisfies $R<\frac{W-\tau }{\delta }$. Then, the tail behavior of the
commuting distance cannot follow a power law if $W$ has a bounded
distribution and we therefore allow $W$ to have an unbounded distribution.
The density of commuting distances is 
\begin{align}
P\left( R=r|W-\delta R>\tau \right) &=\frac{P\left( R=r\right) P\left(
W-\delta r>\tau \right) }{P\left(W-\delta R>\tau \right) }  \notag \\
&=\frac{2\pi \rho r\left( 1-F\left( \tau +\delta r\right) \right) }{%
\int_{0}^{\infty }2\pi \rho s\left( 1-F\left( \tau +\delta s\right) \right)
ds}.
\end{align}
From this, we can observe that 
\begin{equation}
\frac{\partial \ln P}{\partial \ln r}=1-\frac{f\left( \tau +\delta r\right) 
}{1-F\left( \tau +\delta r\right) }\delta r.
\end{equation}%
which shows that in general $P$ does not decay as a power law, unless $\frac{%
f\left( \tau +\delta r\right) }{1-F\left( \tau +\delta r\right) }= \frac{%
\zeta }{r}$ for some $\zeta >1$. In the specific case where $W$ follows a
power law with $F\left( W\right) =1-W^{-\zeta },W>1$, we obtain 
\begin{equation}
\frac{\partial \ln P}{\partial \ln r}=1-\frac{\zeta r}{\tau +\delta r},
\end{equation}%
which tends to $1-\frac{\zeta }{\delta }$ as $r\rightarrow \infty $. This
model thus leads to a power law for the distribution of commute distances,
if the distribution of wage offers follows a power law. If we consider all
wages, the Pareto law tells us that they can be broadly distributed, but
this is not the quantity needed here. Indeed we are considering here the
offer distribution for a given set of skills and it is very unlikely that a
given individual will sample offers that range over the whole income
distribution.

We can then compute the relationship between the average commuting distance
and income in this model. For $w-\delta r>\tau $, we have%
\begin{align}
&P\left( R=r|W=w,W-\delta R>\tau \right) =  \notag \\
&=\frac{P\left( R=r,W=w,w-\delta r>\tau \right) }{P\left( W=w,\delta
R<w-\tau \right) }  \notag \\
&=\frac{2r}{\left( \frac{w-\tau }{\delta }\right) ^{2}},
\end{align}
which leads to the conditional expectation%
\begin{eqnarray}
E\left( R|W=w,W-\delta R>\tau \right) &=&\int_{0}^{\frac{w-\tau }{\delta }}%
\frac{2r^{2}}{\left( \frac{w-\tau }{\delta }\right) ^{2}}\mathrm{d}r  \notag
\\
&=&\frac{2}{3}\frac{w-\tau }{\delta }.
\end{eqnarray}
This model thus predicts that for a linear transport cost, the expected
commute distance is always linear in income which does not fit the empirical
findings.

In any case, it seems that in order to predict results consistent with
empirical observations (a broad law such as a power law with exponent close
to $3$ for the distribution, and a power law behavior for the average
distance), this model needs fine-tuning of the parameters, in sharp contrast
with the closest opportunity model.

\subsection{Including transport costs in the closest opportunity model}

Workers base their decisions on transport costs that depend not only on
distance but also on monetary costs and travel time. We shall see how
transport costs can be accommodated by the closest opportunity model
proposed in this paper. This is useful as we get exact predictions regarding
how the observables of the model are modified by transport costs. The model
can then also be used for prediction in cases when transport costs change.

We let the variable $r$ represent here the transport cost, the closest
opportunity model predicts 
\begin{equation}
\mathrm{d}\log P\left( R=r\right) /\mathrm{d}\log (r)=-3
\end{equation}
In general we may expect that the transport cost is an increasing and
concave function of distance, since travelers switch to faster modes for
longer trips. Denoting the physical distance by $\ell $, we assume that $%
r\sim \ell ^{\nu }$, where $0<\nu <1$. In terms of distance we then find
that 
\begin{equation}
\frac{\mathrm{d}\log P\left( R=r\right) }{\mathrm{d}\log \ell }=\frac{%
\mathrm{d}\log P\left( R=r\right) }{\mathrm{d}\log r}\frac{\mathrm{d}\log r}{%
\mathrm{d}\log \ell }=-3\nu .
\end{equation}%
For the income elasticity, the model predicts a relationship between
transport cost and income that is $\beta =\mathrm{d}\log \overline{r(I)} /%
\mathrm{d}\log I=1/2$. The elasticity of commuting distance with respect to
income is then larger:%
\begin{equation*}
\frac{\mathrm{d}\log \overline{r^{1/\nu}(Y)} }{\mathrm{d}\log Y}=1/2+\frac{1%
}{2\nu }.
\end{equation*}

Observing the commuting distance rather than transport cost, we thus expect
an exponent in the tail of the distribution smaller than 3 in absolute value
and an income elasticity of the average commuting distance that is greater
than 1/2. It is thus possible to back out the exponent $\nu $ from both
observed exponents.

\section{Data Accessibility}

For UK data~\cite{UKdata}, a special licence access can be provided upon
request to the Department for Transport, National Travel Survey, 2002-2012:
Special Licence Access [computer file]. 2nd Edition. Colchester, Essex: UK
Data Archive [distributor], December 2014. SN: 7553 , %
\url{http://dx.doi.org/10.5255/UKDA-SN-7553-2}.

USA data~\cite{USdata} are provided by the U.S. Department of
Transportation, Federal Highway Administration, National Household Travel
Survey. These can by download from URL address: \url{http://nhts.ornl.gov.}

The Danish original dataset is not in open access. Please contact the
authors for further information about the accessibility of aggregated data.



%
%

\section{Authors' contributions}

All authors (GC, IM, MF, MB) participated to the design of the study and
drafted the manuscript. IM prepared the Danish data and GC carried out the
data analysis.


\section{Acknowledgements}

We acknowledge the Departement of Transport, the National Centre for Social
Research, the UK Data Archive and the Crown, for the UK dataset. We
thank Statistics Denmark for providing us with the danish
dataset. Giulia Carra
thanks the Complex Systems Institute in Paris (ISC-PIF) for hosting her
during part of this work. Mogens Fosgerau and Ismir Mulalic have received
funding from the Danish Innovation Fund under the Urban project. Marc
Barthelemy thanks Jean-Philippe Bouchaud, Pablo Jensen, Cl\'ement Sire, and Guy Theraulaz for
stimulating discussions.

\bibliographystyle{prsty}

\end{document}